\documentclass[vecphys]{svmult}

\usepackage{makeidx}         % allows index generation
\usepackage{graphicx}        % standard LaTeX graphics tool
\usepackage{multicol}        % used for the two-column index
\usepackage[bottom]{footmisc}% places footnotes at page bottom
\newcommand{\e}{\mbox{e}}
\newcommand{\im}{\mbox{i}}

\makeindex             % used for the subject index
\begin{document}

\title*{The density-matrix renormalization group applied to transfer matrices:
Static and dynamical properties of one-dimensional quantum systems at finite
temperature}
\titlerunning{DMRG applied to transfer matrices}
\author{Stefan Glocke\inst{1}\and Andreas Kl\"umper\inst{1} \and Jesko Sirker\inst{2}}
\institute{Bergische Universit\"at Wuppertal, Fachbereich Physik, 42097
  Wuppertal, Germany %% \texttt{glocke@physik.uni-wuppertal.de}, \texttt{kluemper@physik.uni-wuppertal.de}
\and Department of Physics and Astronomy, University of British Columbia,
Vancouver, BC, V6T 1Z1, Canada} %% \texttt{sirker@physics.ubc.ca}}
\maketitle

The density-matrix renormalization group (DMRG) applied to transfer matrices
allows it to calculate static as well as dynamical properties of
one-dimensional quantum systems at finite temperature in the thermodynamic
limit. To this end the quantum system is mapped onto a two-dimensional
classical system by a Trotter-Suzuki decomposition. Here we discuss two
different mappings: The standard mapping onto a two-dimensional lattice with
checkerboard structure as well as an alternative mapping introduced by two of
us. For the classical system an appropriate quantum transfer matrix is defined
which is then treated using a DMRG scheme. As applications, the calculation
of thermodynamic properties for a spin-$1/2$ Heisenberg chain in a staggered
magnetic field and the calculation of boundary contributions for open spin
chains are discussed. Finally, we show how to obtain real time dynamics from a
classical system with complex Boltzmann weights and present results for the
autocorrelation function of the $XXZ$-chain.
\section{Introduction}
\label{sec:1}
% Always give a unique label
% and use \ref{<label>} for cross-references
% and \cite{<label>} for bibliographic references
% use \sectionmark{}
% to alter or adjust the section heading in the running head
Several years after the invention of the DMRG method to study ground-state
properties of one-dimensional (1D) quantum systems \cite{WhiteDMRG}, Nishino
showed that the same method can also be applied to the transfer matrix of a
two-dimensional (2D) classical system hence allowing to calculate its
partition function at finite temperature \cite{Nishino}. The same idea can
also be used to calculate the thermodynamic properties of a 1D quantum system
after mapping it to a 2D classical one with the help of a Trotter-Suzuki
decomposition \cite{Trotter,Suzuki1,Suzuki2}.  Bursill {\it et. al.}
\cite{BursillXiang} then presented the first application but the density
matrix chosen in this work to truncate the Hilbert space was not optimal so
that the true potential of this new numerical method was not immediately
clear.  This changed when Wang and Xiang \cite{WangXiang} and Shibata
\cite{Shibata} presented an improved algorithm and showed that the
density-matrix renormalization group applied to transfer matrices (which we
will denote as TMRG from hereon) is indeed a serious competitor to other
numerical methods as for example Quantum-Monte-Carlo (QMC). Since then, the
TMRG method has been successfully applied to a number of systems including
various spin chains, the Kondo lattice model, the $t-J$ chain and ladder and
also spin-orbital models
\cite{EggertRommer,Raupach,RiceTroyer,SirkerKhaliullin,SirkerSu4,MutouShibata,SirkerKluemperEPL,SirkerKluemperPRB,NaefWang}
.

The main advantage of the TMRG algorithm is that the thermodynamic limit can
be performed exactly thus avoiding an extrapolation in system size.
Furthermore, there are no statistical errors and results can be obtained with
an accuracy comparable to $T=0$ DMRG calculations. Similar to the $T=0$ DMRG
algorithms, the method is best suited for 1D systems with short range
interactions. These systems can, however, be either bosonic or fermionic
because no negative sign problem as in QMC exists. Most important, there are
two areas where TMRG seems to have an edge over any other numerical methods
known today.  These are: (1) Impurity or boundary contributions, and (2)
real-time dynamics at finite temperature. As first shown by Rommer and Eggert
\cite{RommerEggert}, the TMRG method allows it to separate an impurity or
boundary contribution from the bulk part thus giving direct access to
quantities which are of order $\mathcal{O}(1/L)$ compared to the
$\mathcal{O}(1)$ bulk contribution (here $L$ denotes the length of the
system). We will discuss this in more detail in section \ref{Imp}. Calculating
numerically the dynamical properties for large or even infinite 1D quantum
systems constitutes a particularly difficult problem because QMC and TMRG
algorithms can usually only deal with imaginary-time correlation functions.
The analytical continuation of numerical data is, however, an ill-posed
problem putting severe constraints on the reliability of results obtained this
way.  Very recently, two of us have presented a modified TMRG algorithm which
allows for the direct calculation of real-time correlations
\cite{SirkerKluemperDTMRG}.  This new algorithm will be discussed in section
\ref{RealTime}.

Before coming to these more recent developments we will discuss the definition
of an appropriate quantum transfer matrix for the classical system in section
\ref{QTM} and describe how the DMRG algorithm is applied to this object in
section \ref{DMRGAlgo}. Here we will follow in parts the article by Wang and Xiang
in \cite{Peschel} but, at the same time, also discuss an alternative
Trotter-Suzuki decomposition \cite{SirkerKluemperEPL,SirkerKluemperPRB}. 
\section{Quantum transfer matrix theory}
\label{QTM}
\begin{figure}
%%\centering
\includegraphics[width=0.99\columnwidth]{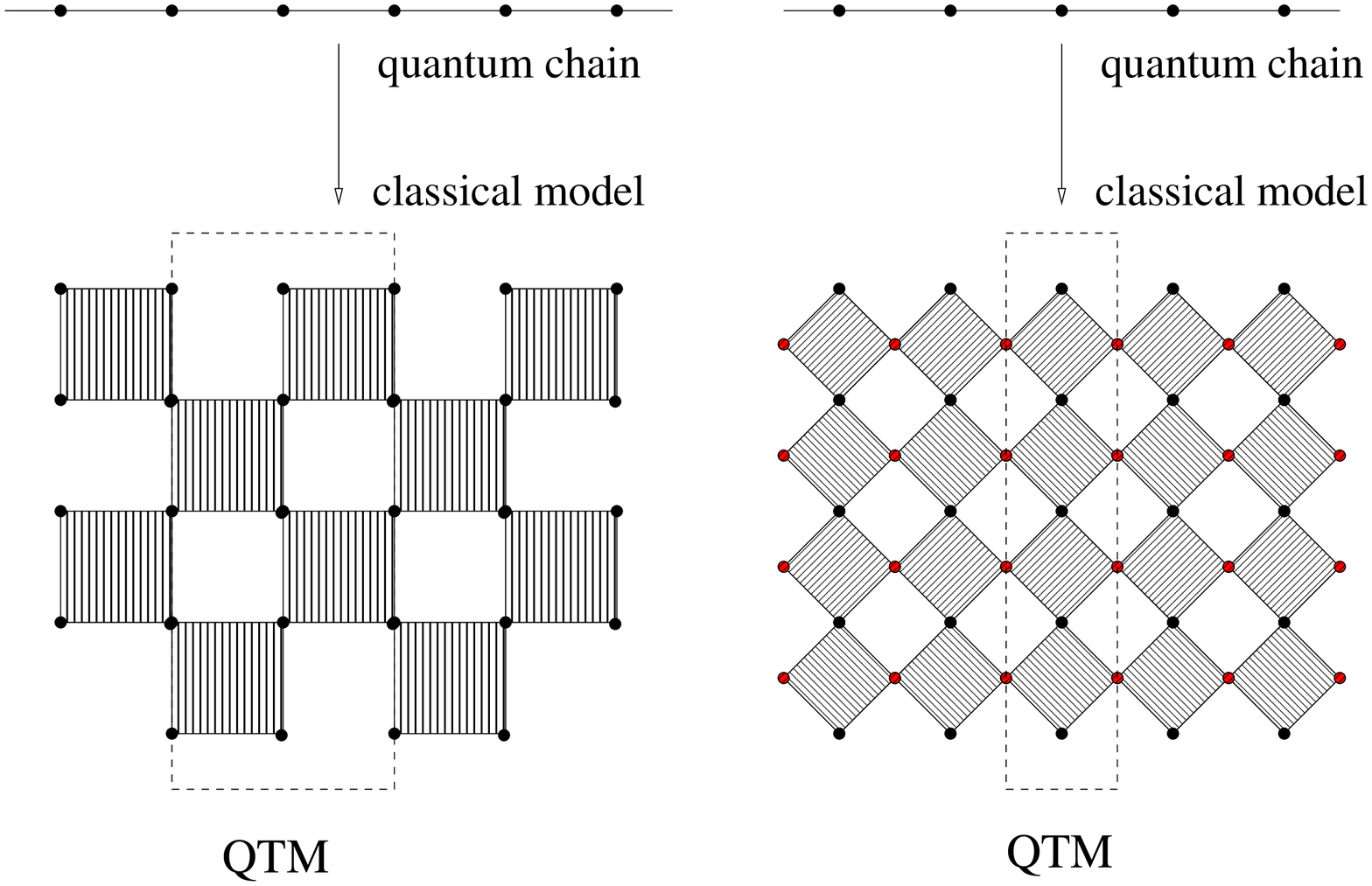}
\caption{The left part shows the standard Trotter-Suzuki mapping of the 1D quantum chain to a 2D
classical model with checkerboard structure where the vertical direction corresponds to imaginary
time. The QTM is two-column wide. The right part shows the
alternative mapping. Here, the QTM is only one column wide.}
\label{fig:mapping}    
\end{figure}

The TMRG method is based on a Trotter-Suzuki decomposition of the partition
function, mapping a 1D quantum system to a 2D classical one
\cite{Trotter,Suzuki1,Suzuki2}. In the following, we discuss both the standard
mapping introduced by Suzuki \cite{Suzuki2} as well as an alternative one
\cite{SirkerKluemperEPL,SirkerKluemperPRB} starting from an arbitrary
Hamiltonian $H$ of a 1D quantum system with length $L$, periodic boundary
conditions and nearest neighbor interaction
\begin{equation}
 H = \sum_{i=1}^{L} h_{i,i+1}\;.
\end{equation}
The standard mapping, widely used in QMC and TMRG calculations, is described
in detail in \cite{Peschel}. Therefore we only summarize it briefly here.
First, the Hamiltonian is decomposed into two parts, $H = H_e + H_o$, where
each part is a sum of commuting terms.
%% \begin{equation}
%% \nonumber
%% H_{o(e)} = \sum_{i=\mbox{odd(even)}} h_{i,i+1} \;.
%% \end{equation}
Here $H_e$ ($H_o$) contains the interactions $h_{i,i+1}$ with $i$ even (odd).
By discretizing the imaginary time, the partition function becomes
\begin{equation}
\label{eqn_part_func_trad}
Z = \mbox{Tr}\;\mbox{e}^{-\beta H}=\lim_{M\to\infty}\mbox{Tr}\left\{\left[\mbox{e}^{-\epsilon H_e}\mbox{e}^{-\epsilon H_o}\right]^M\right\}
\end{equation}
with $\epsilon=\beta/M$, $\beta$ being the inverse temperature and $M$ an
integer (the so called Trotter number). By inserting $2M$ times a
representation of the identity operator, the partition function is expressed by
a product of local Boltzmann weights
\begin{equation}
\tau_{k,k+1}^{i,i+1}=\left<s_k^i s_k^{i+1}\right|\mbox{e}^{-\epsilon H_{e,o}}\left|s_{k+1}^i s_{k+1}^{i+1}\right>\;,
\end{equation}
denoted in a graphical language by a shaded plaquette (see
Fig.~\ref{fig:mapping}). The subscripts $i$ and $k$ represent the spin
coordinates in the space and the Trotter (imaginary time) directions,
respectively. A column-to-column transfer matrix $\mathcal{T}_M$, the so called {\it
  quantum transfer matrix} (QTM), can now be defined using these local
Boltzmann weights
\begin{equation}
 \mathcal{T}_M = \left(\tau_{1,2} \tau_{3,4}\ldots\tau_{2M-1,2M}\right)\left(\tau_{2,3}
   \tau_{4,5}\ldots\tau_{2M,1}\right) \;.
\end{equation}
and is shown in the left part of Fig.~\ref{fig:mapping}. The partition
function is then simply given by
\begin{equation}
Z=\mbox{Tr}\;\mathcal{T}_M^{L/2}\; .  %=\sum_\mu \Lambda_\mu^{L/2} \; .
\end{equation}
The disadvantage of this Trotter-Suzuki mapping to a 2D lattice
with checkerboard structure is that the QTM is two columns wide. This
increases the amount of memory necessary to store it and also complicates the
calculation of correlation functions. 

Alternatively, the partition function can also be expressed by \cite{SirkerKluemperEPL,SirkerKluemperPRB}
\begin{equation}
Z = \lim_{M\to\infty} \mbox{Tr}
\left\{\left[\mathcal{T}_1(\epsilon)\mathcal{T}_2(\epsilon)\right]^{M/2}\right\}\; ,
\end{equation}
with $\mathcal{T}_{1,2}(\epsilon)=T_{R,L}\exp[-\epsilon H + \mathcal{O}(\epsilon^2)]$.
Here, $T_{R,L}$ are the right- and left-shift operators, respectively. The obtained
classical lattice has alternating rows and additional points in a mathematical
auxiliary space. Its main advantage is that it allows to formulate a QTM which is only one column wide
(see right part of Fig.~\ref{fig:mapping}). 
The derivation of this QTM is completely analogous to the standard one, even
the shaded plaquettes denote the same Boltzmann weight. Here, however, these
weights are rotated by 45$^0$ clockwise and anti-clockwise in an alternating
fashion from row to row. Using this transfer matrix,
$\widetilde{\mathcal{T}}_M$, the partition function is given by
$Z=\mbox{Tr}\;\widetilde{\mathcal{T}}_M^{L}$.
%% The partition
%% function is given by
%% \begin{equation}
%% Z=\mbox{Tr}\;T_M^{L}\;.%=\sum_\mu \Lambda_\mu^{L} \;.
%% \end{equation}
%% In the following
%% we will always use the novel mapping, since the traditional mapping is exactly
%% described in \cite{DMRG_book}. 
\subsection{Physical properties in the thermodynamic limit}
The reason why this transfer matrix formalism is extremely useful for
numerical calculations has to do with the eigenspectrum of the QTM. At
infinite temperature it is easy to show \cite{SirkerDiss} that the largest
eigenvalue of the QTM $\mathcal{T}_M$ ($\widetilde{\mathcal{T}}_M$) is given
by $S^2$ ($S$) and all other eigenvalues are zero. Here $S$ denotes the number
of degrees of freedom of the physical system per lattice site. Decreasing the
temperature, the gap between the leading eigenvalue $\Lambda_0$ and
next-leading eigenvalues $\Lambda_n$ ($n>0$) of the transfer matrix shrinks.
The ratio between $\Lambda_0$ and each of the other eigenvalues $\Lambda_n$,
however, defines a {\it correlation length} $1/\xi_n
=\ln|\Lambda_0/\Lambda_n|$ \cite{Peschel,SirkerDiss}. Because a 1D quantum
system cannot order at finite temperature, any correlation length $\xi_n$ will
stay finite for $T>0$, i.e., the gap between the leading and any next-leading
eigenvalue stays finite. Therefore the calculation of the free energy in the
{\it thermodynamic limit} boils down to the calculation of the largest
eigenvalue $\Lambda_0$ of the QTM
\begin{eqnarray}
 \nonumber
f&=&-\lim_{L\to\infty}\frac 1 {\beta L} \ln Z =
-\lim_{L\to\infty}\lim_{\epsilon\to 0}\frac 1{\beta L}  \ln\mbox{Tr}\, \widetilde{\mathcal{T}}_M^L\\
&=&  -\lim_{\epsilon\to 0}\lim_{L\to \infty}\frac 1{\beta L}\ln\bigg\{\Lambda_0^L\bigg[1+\sum_{l>1}
\underbrace{\left(\Lambda_l/\Lambda_0 \right)^L}_{\stackrel{L\to \infty}{\longrightarrow 0}}\bigg]\bigg\} \nonumber\\
&=& -\lim_{\epsilon\to 0}\frac{\ln\Lambda_0}{\beta}\; .
\end{eqnarray}
Here the interchangeability of the limits $L\rightarrow \infty$ and
$\epsilon\rightarrow 0$ has been used \cite{Suzuki2}. Local expectation values
and static two-point correlation functions can be calculated in a similar
fashion (see e.g.~\cite{Peschel} and \cite{SirkerDiss}).
%%%%%%%%%%%%neu von Herrn Klümper%%%%%%%%%%%%%%
In the next section, we are going to show how the eigenvalues of the QTM are
computed by means of the density matrix renormalization group. This is
possible since the transfer matrices are built from local objects. Instead of
sums of local objects we are dealing with products, but this is not essential
to the numerical method. However, there are a few important differences in
treating transfer matrices instead of Hamiltonians. At first sight, these
differences look technical, but at closer inspection they reveal a physical
core.

The QTMs as introduced above are real valued, but not symmetric. This is not a
serious drawback for numerical computations, but certainly inconvenient. So
the first question that arises is whether the transfer matrices can be
symmetrized. Unfortunately, this is not the case. If the transfer matrix were
replaceable by a real symmetric (or a hermitean) matrix all eigenvalues would
be real and the ratios of next-leading eigenvalues to the leading eigenvalue
would be real, positive or negative.  Hence all correlation functions would
show commensurability with the lattice.  However, we know that a generic
quantum system at sufficiently low temperatures yields incommensurate
oscillations with wave vectors being multiples of the Fermi vector taking
rather arbitrary values.

Therefore we know that the spectrum of a QTM must consist of real eigenvalues
or of complex eigenvalues organized in complex conjugate pairs. This opens the
possibility to understand the QTM as a {\it normal matrix} upon a suitable
choice of the underlying scalar product. Unfortunately, the above introduced
matrices are not {\it normal} with respect to standard scalar products, i.e.
we do not have
$[\widetilde{\mathcal{T}}_M,\widetilde{\mathcal{T}}_M^\dagger]=0$.
%%%%%%%%%%%%%%%%%%%%%%%%%%%%%%%%%%%%%%%%%%%%%%%%%%%%%%%%%%%%%%%%%%%%%
%%%%%%%%%%%%%%%%%%%%%%%%%%%%%%%%%%%%%%%%%%%%%%%%%%%%%%%%%%%%%%%%%%
%%%%%%%%%%%%%%%%%%%%%%%%%%%%%%%%%%%%%%%%%%%%%%%%%%%%%%%%%%%%%%%%%%
\section{The Method - DMRG algorithm for the QTM}
\label{DMRGAlgo}
Next, we describe how to increase the length of the transfer matrix in
imaginary time, i.e. the inverse temperature, by successive DMRG steps.  Like
in the ordinary DMRG, we first divide the QTM into two parts, the system $S$
and the environment block $E$. Using the QTM, $\widetilde{\mathcal{T}}_M$, the
density matrix is defined by
\begin{equation}
  \rho=\widetilde{\mathcal{T}}_M^L\;,
\end{equation}
which reduces to $\rho=\left|\Psi_0^R\right>\left<\Psi_0^L\right|$ up to a
normalization constant in the thermodynamic limit. As in the zero temperature
DMRG algorithm, a reduced density matrix $\rho_S$ is obtained by taking a
partial trace over the environment
\begin{equation}
  \rho_S = \mbox{Tr}_E\{\left|\Psi_0^R\right>\left<\Psi_0^L\right|\} \; .
\end{equation}
Note that this matrix is real but non-symmetric, which complicates its numerical
diagonalization. It also allows for complex conjugated pairs of eigenvalues
which have to be treated separately (see \cite{SirkerDiss} for details).

In actual computations, the Trotter-Suzuki parameter $\epsilon$ is fixed.
Therefore the temperature $T\sim 1/\epsilon M$ is decreased by an iterative
algorithm $M\to M+1$.  In the following, the blocks of the QTM,
$\widetilde{\mathcal{T}}_M$, are shown in a $90^\circ$-rotated view.
%% Note that in all steps below the cases $M$ even and
%% odd have to be treated separately.
\begin{enumerate}
\item First we construct the initial system block $\Gamma$ consisting of $M$
  plaquettes so that $S^M \le N < S^{M+1}$, where $S$ is the dimension of the
  local Hilbert space and $N$ is the number of states which we want to keep.
\begin{figure}
\centering
\includegraphics[height=2.5cm]{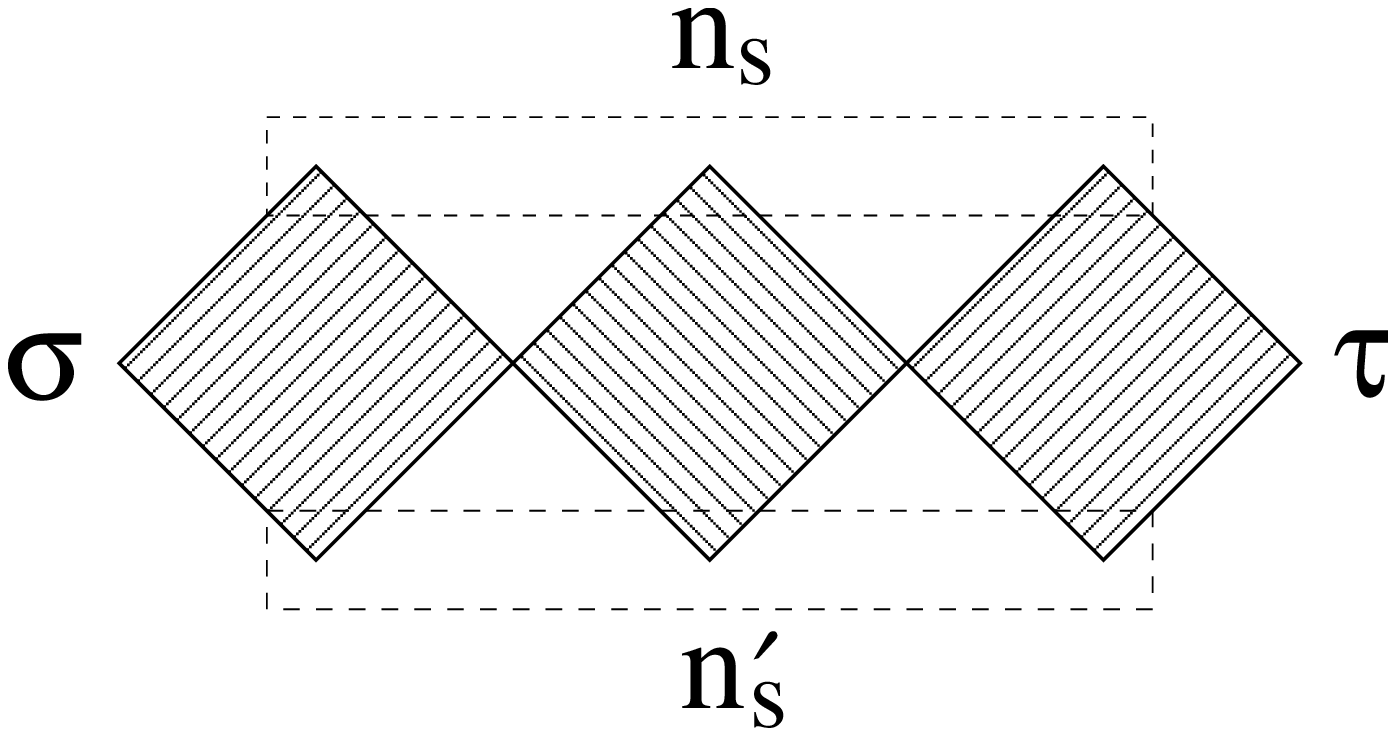}
\caption{The system block $\Gamma$. The plaquettes are connected by a
  summation over the adjacent corner spins.}
\label{fig:algo1}    
\end{figure}
$n_s,n_s^\prime$ are block-spin variables and contain
$\widetilde{N}=S^M$ states. The $S^2\cdot \widetilde{N}^2$-dimensional array
$\Gamma(\sigma,n_s,\tau,n_s^\prime)$ is stored.
\item The enlarged system block
  $\widetilde\Gamma(\sigma,n_s,s_2,\tau,s_2^\prime,n_s^\prime)$, a $S^4\cdot \widetilde
  N^2$-dimensional array, is formed by adding a plaquette to the system block.
  If $h_{i,i+1}$ is real and translationally invariant, the environment block
  can be constructed by a $180^\circ$-rotation and a following inversion of
  the system block. Otherwise the environment block has to be treated
  separately like the system block. Together both blocks form the superblock
  (see Fig. \ref{fig:algo2}). 
\begin{figure}
\centering
\includegraphics[width=0.99\columnwidth]{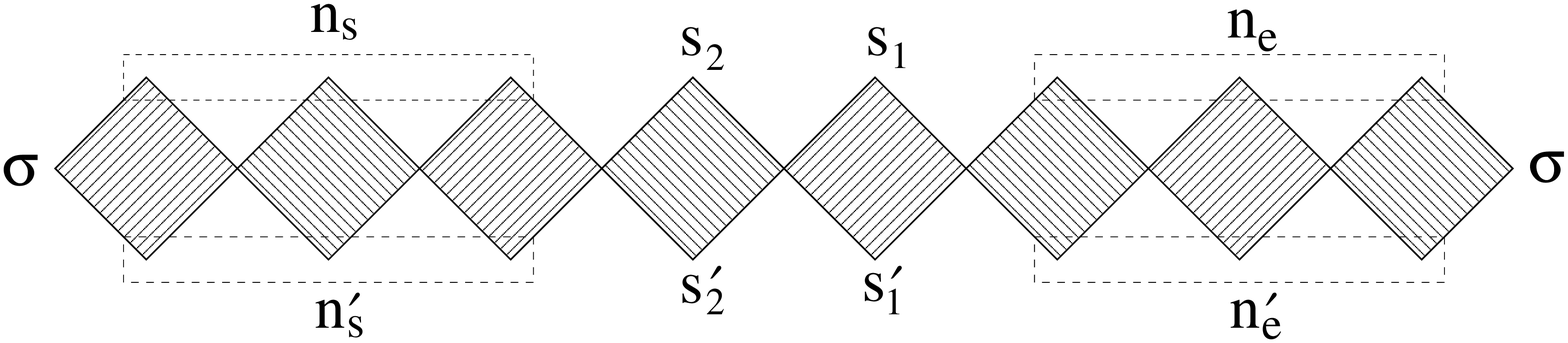}
\caption{The superblock is closed periodically by a summation over all
  $\sigma$ states. }
\label{fig:algo2}    
\end{figure}
\item The leading eigenvalue $\Lambda_0$ and the corresponding left and right
  eigenstates 
\begin{equation}
\left<\Psi_0^L\right|=\Psi^L(s_1,n_s,s_2,n_e)\;,\;\left|\Psi_0^R\right>=\Psi^R(s_1^\prime,n_s^\prime,s_2^\prime,n_e^\prime) 
\end{equation}
are calculated and normalized $\langle\Psi_0^L|\Psi_0^R\rangle =1$. Now
thermodynamic quantities can be evaluated at the temperature
$T=1/(2\epsilon(M+1))$.
%\item At the temperature $T=1/2\epsilon(M+1)$ the thermodynamic quantities
%are evaluated. 
\item A reduced density matrix is calculated by performing the trace over the
  environment
\begin{eqnarray}
\rho_s(n_s^\prime,s^\prime_2|n_s,s_2)&=&\sum_{s1,n_e}\left|\Psi_0^R\right>\left<\Psi_0^L\right| \\
&=&\sum_{s1,n_e}\Psi^R(s_1,n_s^\prime,s_2^\prime,n_e)\Psi^L(s_1,n_s,s_2,n_e) \nonumber
\end{eqnarray}
and the complete spectrum is computed. A $N\times (S\cdot\widetilde N)$-matrix
$V^L(\widetilde n_s|n_s,s_2)$ ($V^R(\widetilde n_s^\prime|n_s^\prime,s_2^\prime)$) is
constructed using the left (right) eigenstates belonging to the $N$ largest
eigenvalues, where $\widetilde n_s$ ($\widetilde n_s^\prime$) is a new renormalized
block-spin variable with only $N$ possible values.
\item Using $V^L$ and $V^R$ the system block is renormalized.
\begin{figure}
\centering
\includegraphics[width=0.99\columnwidth]{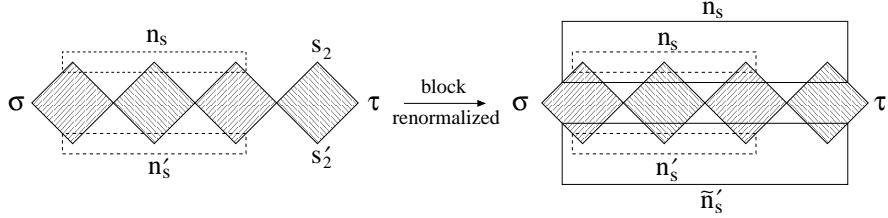}
\caption{The renormalization step for the system block.}
\label{fig:algo3}  
\end{figure}
The renormalization is given by
\begin{equation}
  \Gamma(\sigma,\widetilde n_s,\tau,\widetilde
  n_s^\prime)=\sum_{n_s,s_2}\sum_{n_s^\prime,s_2^\prime}V^L(\widetilde n_s|n_s,s_2)\widetilde\Gamma(\sigma,n_s,s_2,\tau,s_2^\prime,n_s^\prime)V^R(\widetilde n_s^\prime|n_s^\prime,s_2^\prime)\;.
\end{equation}
\end{enumerate}
Now the algorithm is repeated starting with step 2 using the new system block.
However, the block-spin variables can now take $N$ instead of $\widetilde N$
values.
\section{An example: The spin-$1/2$ Heisenberg chain with staggered and
  uniform magnetic fields} 
\label{sec:3:CuPM}
\begin{figure}
\centering
\includegraphics*[width=0.99\columnwidth]{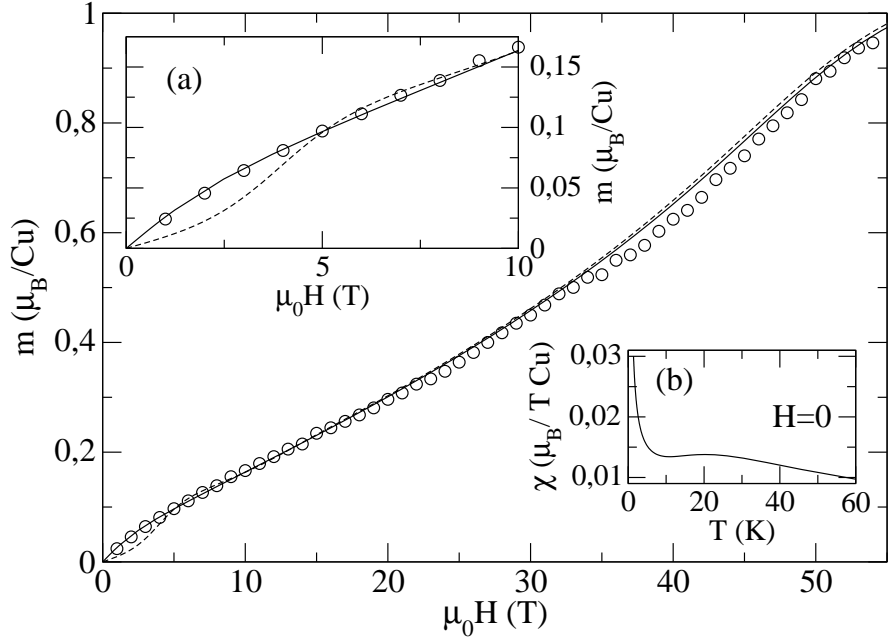}%{CuPM_0c11}
\caption{TMRG data (solid line) and experimental magnetization curves
  (circles) for CuPM at a temperature $T = 1.6$~K with the magnetic field
  applied along the $c^{\prime\prime}$~axis. For comparison ED data for a
  system of $16$ sites and $T=0$ are shown (dashed lines). Here $J/k_B =
  36.5$~K, $h_u = g\mu_B H/J$, $h_s = 0.11\, h_u$ and $g=2.19$. Inset~(a):
  Magnetization for small magnetic fields. Inset~(b): Susceptibility as a
  function of temperature $T$ at $H=0$ calculated by TMRG.}
\label{fig:CuPM}
\end{figure}

As example, we show here results for the magnetization of a spin-$1/2$
Heisenberg chain subject to a staggered magnetic field $h_s$ and a uniform
field $h_u=g\mu_B H/J$
\begin{equation}
\mathcal{H}= J \sum_i\left[\mathbf{S}_i\mathbf{S}_{i+1} - h_u S_i^z - (-1)^i h_s
  S_i^x\right] \; ,
\end{equation}
where $H$ is the external uniform magnetic field and $g$ the Land\'e factor.
An effective staggered magnetic field is realized in spin-chain compounds as
for example copper pyrimidine dinitrate (CuPM) or copper benzoate if an
external uniform magnetic field $H$ is applied \cite{OshikawaAffleck}. For CuPM
the magnetization as a function of applied magnetic field $H$ has been measured
experimentally. In Fig. \ref{fig:CuPM} the excellent agreement between these
experimental and TMRG data at a temperature $T= 1.6$~K with a magnetic field
applied along the $c^{\prime\prime}$ axis is shown. Along the $c^{\prime\prime}$ axis the effect
due to the induced staggered field is largest (see
\cite{Glocke} for more details). Note that at low magnetic fields the TMRG data
describe the experiment more accurately than the exact diagonalization (ED)
data, because there are no finite size effects (see inset (a) of Fig. \ref{fig:CuPM}).
%%%%%%%%%%%%%%%%%%%%%%%%%%%%%%%%%%%%%%%%%%%%
%% Questions:
%% 1) How many sites in ED?
%% 2) 
%\begin{figure}
%\begin{minipage}[c]{0.5\textwidth}
%\centering 
%\includegraphics*[width=0.95\textwidth]{CuPM_0c11_suscep}%{CuPM_0c11}
%\caption{Theoretical and experimental magnetization curves for CuPM at a
%  temperature $T = 1.6$~K along the $c^{\prime\prime}$~axis. Solid (dashed)
%  line: TMRG (ED) data and open circles: experimental data. Inset~(a):
%  Magnetization for small magnetic fields. Inset~(b): Susceptibility as a
%  function of temperature $T$ at $H=0$ calculated by TMRG. Here $J/k_B = 36.5$~K, $h_u =
%  g\mu_B H/J$, $h_s = 0.11\, h_u$ and $g=2.19$.}
%\label{fig:CuPM}
%\end{minipage}%
%\begin{minipage}[c]{0.5\textwidth}
%\centering \includegraphics[width=0.95\textwidth]{suszep_cupm} \caption{Gro{\ss}e Box}
%\label{fig:big}
%\end{minipage} 
%\end{figure}
For a magnetic field $H$ applied along the $c^{\prime\prime}$ axis a gap,
$\Delta \propto H^{2/3}$, is induced with multiplicative logarithmic
corrections. For $H\to 0$ and low $T$ the susceptibility diverges $\chi\sim
1/T$ because of the staggered part \cite{OshikawaAffleckCUB} (see inset (b) of
Fig. \ref{fig:CuPM}).
%\begin{figure}
%\centering
%\includegraphics*[height=5.5cm]{suszep_cupm}
%\caption{asdsad}
%\label{fig:CuPM_suscep}
%\end{figure}
\section{Impurity and boundary contributions}
\label{Imp}
In recent years much interest has focused on the question how impurities and
boundaries influence the physical properties of spin chains
\cite{EggertAffleck92,FujimotoEggert,SirkerBortzJSTAT,FurusakiHikihara,SirkerLaflorencie}.
The doping level $p$ defines an average chain length $\bar{L}=1/p-1$ and
impurity or boundary contributions are of order $\sim \mathcal{O}(1/\bar{L})$
compared to the bulk.  This makes it very difficult to separate these
contributions from finite size corrections if numerical data for finite
systems (e.g.~from QMC calculations) are used. TMRG, on the other hand, allows
to study directly impurities embedded into an infinite chain
\cite{RommerEggert}. We will discuss here only the simplest case that a single
bond or a single site is different from the rest. The partition function is
then given by
\begin{equation}
  Z=\mbox{Tr}\left(\widetilde{\mathcal{T}}_M^{L-1}\mathcal{T}_{imp}\right)\;,
\end{equation}
where $\mathcal{T}_{imp}$ is the QTM describing the site impurity or the modified
bond. In the thermodynamic limit the total free energy then becomes
\begin{equation}
F=-T\ln Z = Lf_{\mbox{\tiny bulk}} + F_{\mbox{\tiny
    imp}} = -LT\ln\Lambda_0 -T\ln(\lambda_{imp}/\Lambda_0) \; , 
\end{equation}
with $\Lambda_0$ being the largest eigenvalue of the QTM, $\widetilde{\mathcal{T}}_M$, and
$\lambda_{imp}=\langle\Psi^L_0|\mathcal{T}_{imp}|\Psi^R_0\rangle$.  

As example, we want to consider a semi-infinite spin-$1/2$ $XXZ$-chain
with an open boundary. In this case translational invariance is broken and
field theory predicts Friedel-type oscillations in the local magnetization
$\langle S^z(r)\rangle$ and susceptibility $\chi(r) = \partial \langle
S^z(r)\rangle /\partial h$ near the boundary
\cite{EggertAffleck95,BortzSirker}. Using the TMRG method the local
magnetization can be calculated by
\begin{equation}
\label{eq_FT5}
\langle S^z(r)\rangle = \frac{\langle\Psi_L^0|\widetilde{\mathcal{T}}(S^z)\widetilde{\mathcal{T}}^{r-1}\mathcal{T}_{\mbox{\tiny
    imp}}|\Psi_R^0\rangle}
{\Lambda_0^r \lambda_{imp}} \; ,
\end{equation}
where $\widetilde{\mathcal{T}}(S^z)$ is the transfer matrix with the operator
$S^z$ included and $\mathcal{T}_{\mbox{\tiny imp}}$ is the transfer matrix
corresponding to the bond with zero exchange coupling. Hence
$\mathcal{T}_{\mbox{\tiny imp}}|\Psi_R^0\rangle$ is nothing but the state
describing the open boundary at the right. 
\begin{figure}[!htp]
\centering
\includegraphics*[width=0.99\columnwidth]{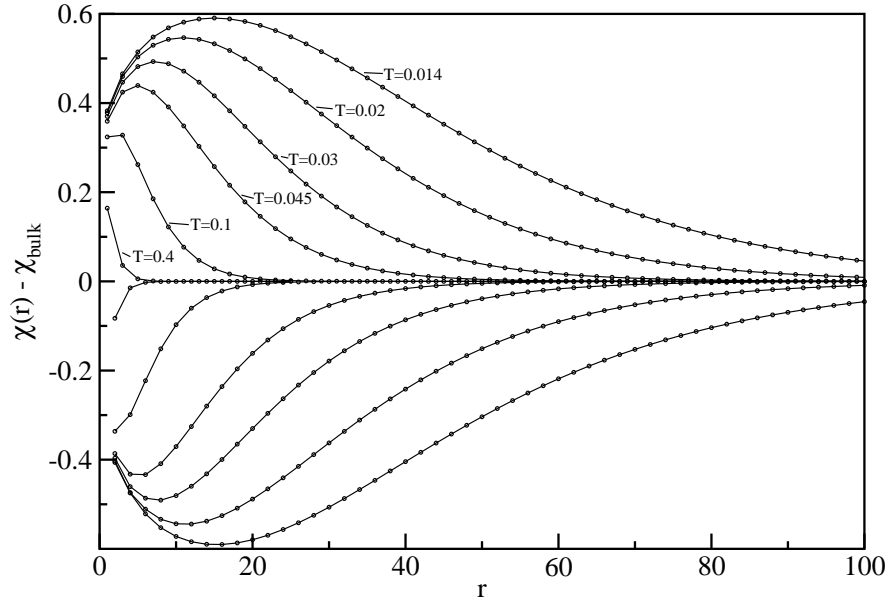}
\caption{Susceptibility profile for $\Delta=0.6$ and different temperatures $T$. $N=240$ states have been
  kept in the DMRG algorithm. The lines are a guide to the eye.}
\label{fig:susciprofile}
\end{figure}
In Fig.~\ref{fig:susciprofile} the susceptibility profile as a
function of the distance $r$ from the boundary for various temperatures as
obtained by TMRG calculations \cite{BortzSirker} is shown.
%\begin{figure}[!htp]
%\begin{center}
%\includegraphics*[width=0.7\columnwidth]{Susci_profile_Delta0.6_different_T.eps}
%\caption{Susceptibility profile for $\Delta=0.6$. $N=240$ states have been
%  kept in the DMRG algorithm. The lines are a guide to the eye.}
%\end{center}
%\label{Fig_susciprofile}
%\end{figure}
For more details the reader is referred to \cite{RommerEggert} and \cite{BortzSirker}.

\section{Real time dynamics}
\label{RealTime}
Finally, we want to discuss a very recent development in the TMRG method. The
Trotter-Suzuki decomposition of a 1D quantum system yields a 2D classical
model with one axis corresponding to imaginary time (inverse temperature). It
is therefore straightforward to calculate imaginary-time correlation functions
(CFs). Although the results for the imaginary-time CFs obtained by TMRG are
very accurate, the results for real-times (real-frequencies) involve errors of
unknown magnitude because the analytical continuation poses an ill-conditioned
problem.  In practice, the maximum entropy method is the most efficient way to
obtain spectral functions from TMRG data. The combination of TMRG and maximum
entropy has been used to calculate spectral functions for the $XXZ$-chain
\cite{NaefWang} and the Kondo-lattice model \cite{MutouShibata}.  However, it
is in principle impossible to say how reliable these results are because of
the afore mentioned problems connected with the analytical continuation of
numerical data. It is therefore desirable to avoid this step completely and to
calculate real-time correlation functions directly.

A TMRG algorithm to do this has recently been proposed by two of us
\cite{SirkerKluemperDTMRG}. Starting point is an arbitrary two-point CF for an
operator $\hat{O}_r(t)$ at site $r$ and time $t$
\begin{equation}
\label{eq1}
\langle\hat{O}_r(t)\hat{O}_0(0)\rangle =
\frac{\mbox{Tr}\left(\hat{O}_r(t)\hat{O}_0(0)\e^{-\beta
      H}\right)}{\mbox{Tr}\left(\e^{-\beta H}\right)} 
= \frac{\mbox{Tr}\left(\e^{-\beta
      H/2}\e^{i t H}\hat{O}_r\e^{-i t H}\hat{O}_0\e^{-\beta
      H/2}\right)}{\mbox{Tr}\left(\e^{-\beta
      H/2}\e^{i t H}\e^{-i tH}\e^{-\beta
      H/2}\right)} \; . 
\end{equation} 
Here we have used the cyclic invariance of the trace and have written the
denominator in analogy to the numerator. In the following we will use the
standard Trotter-Suzuki decomposition leading to a two-dimensional
checkerboard model. 

\begin{figure}[!htp]
\begin{center}
\includegraphics*[width=0.99\columnwidth]{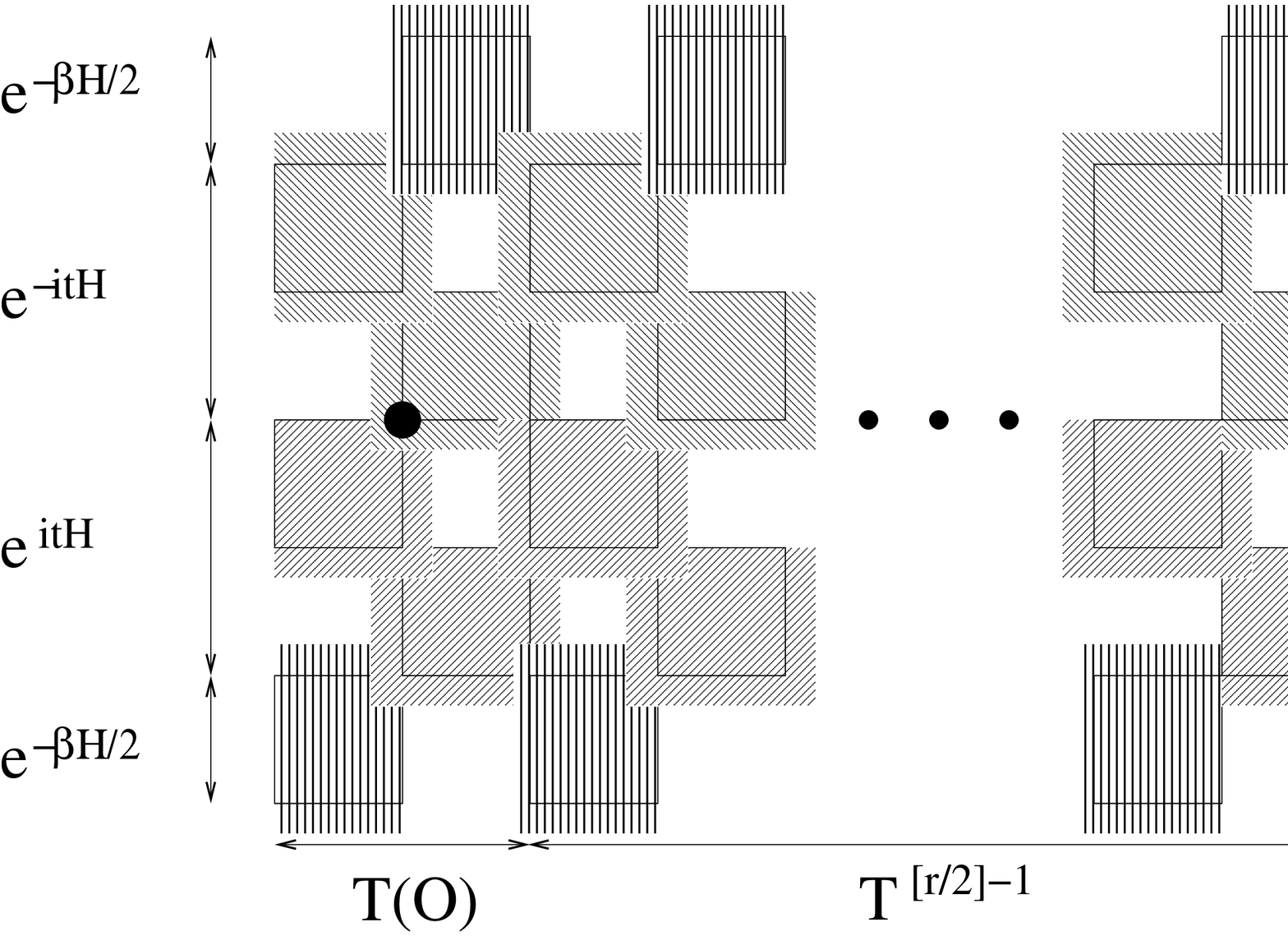}
\end{center}
\caption{Transfer matrices appearing in the numerator of Eq.~(\ref{eq6}) for $r>1$
  with $r$ even. The 2 black dots denote the operator $\mathcal{O}$.
  $\mathcal{T},\mathcal{T}(O)$ consist of three parts: A part representing
  $\exp(-\beta H)$ (vertically striped plaquettes), another for $\exp(\im t
  H)$ (stripes from lower left to upper right) and a third part describing
  $\exp(-\im t H)$ (upper left to lower right). $\mathcal{T},\mathcal{T}(O)$
  are split into system ($S$) and environment ($E$).}
\label{fig1}
\end{figure}

The crucial step in our approach to calculate real-time dynamics directly is
to introduce a second Trotter-Suzuki decomposition of $\exp(-\im\delta H)$
with $\delta=t/N$ in addition to the usual one for the partition function
described in section \ref{QTM}.  We can then define a column-to-column
transfer matrix
\begin{eqnarray}
\label{eq4}
&& \mathcal{T}_{2N,M} =
(\tau_{1,2}\tau_{3,4}\cdots\tau_{2M-1,2M})(\tau_{2,3}\tau_{4,5}\cdots\tau_{2M,2M+1})
 \\
&&
(\bar{v}_{2M+1,2M+2}\cdots\bar{v}_{2M+2N-1,2M+2N}) 
 (\bar{v}_{2M+2,2M+3}\cdots\bar{v}_{2M+2N,2M+2N+1})
\nonumber \\
&& (v_{2M+2N+1,2M+2N+2}\cdots v_{2M+4N-1,2M+4N}) 
 (v_{2M+2N+2,2M+2N+3}\cdots v_{2M+4N,1}) \nonumber
\end{eqnarray}
where the local transfer matrices have matrix elements
\begin{eqnarray}
\label{eq5}
\tau(s_k^i s_k^{i+1}| s_{k+1}^i s_{k+1}^{i+1}) &=& \langle s_k^i s_k^{i+1} | \e^{-\epsilon h_{i,i+1}} |
  s_{k+1}^i s_{k+1}^{i+1} \rangle \\
v(s_l^i s_l^{i+1}| s_{l+1}^i s_{l+1}^{i+1}) &=& \langle s_l^i s_l^{i+1} | \e^{-i\delta h_{i,i+1}} |
  s_{l+1}^i s_{l+1}^{i+1} \rangle \;  \nonumber 
\end{eqnarray}
and $\bar{v}$ is the complex conjugate. Here $i=1,\cdots,L$ is the lattice
site, $k=1,\cdots,2M$ ($l=1,\cdots,2N$) the index of the imaginary time (real
time) slices and $s_{k(l)}^i$ denotes a local basis. The denominator in
Eq.~(\ref{eq1}) can then be represented by
$\mbox{Tr}(\mathcal{T}_{2N,M}^{L/2})$ where $N,M,L\rightarrow\infty$. A
similar path-integral representation holds for the numerator in (\ref{eq1}).
Here we have to introduce an additional modified transfer matrix
$\mathcal{T}_{2N,M}(\hat{O})$ which contains the operator $\hat{O}$ at the
appropriate position. For $r>1$ we find
\begin{eqnarray}
\label{eq6}
\langle\hat{O}_r(t)\hat{O}_0(0)\rangle  &=&\lim_{N,M\rightarrow\infty}\lim_{L\rightarrow\infty}\frac{\mbox{Tr}(\mathcal{T}(\hat{O})\mathcal{T}^{[r/2]-1}\mathcal{T}(\hat{O})\mathcal{T}^{L/2-[r/2]-1})}{\mbox{Tr}(\mathcal{T}^{L/2})}
 \nonumber \\
&=& \lim_{N,M\rightarrow\infty}
\frac{\langle\Psi^L_0|\mathcal{T}(\hat{O})\mathcal{T}^{[r/2]-1}\mathcal{T}(\hat{O})|\Psi^R_0\rangle}{\Lambda_0^{[r/2]+1}\langle\Psi^L_0|\Psi^R_0\rangle}
 \; . 
\end{eqnarray}   
Here $[r/2]$ denotes the first integer smaller than or equal to $r/2$ and we
have set $\mathcal{T}\equiv \mathcal{T}_{2N,M}$. A graphical representation of
the transfer matrices appearing in the numerator of Eq.~(\ref{eq6}) is shown
in Fig.~\ref{fig1}.
This new transfer matrix can again be treated with the DMRG algorithm
described in section \ref{DMRGAlgo} where either a $\tau$ or $v$ plaquette is
added corresponding to a decrease in temperature $T$ or an increase in
real-time $t$, respectively.  

\begin{figure}[!htp]
\begin{center}
\includegraphics*[width=0.99\columnwidth]{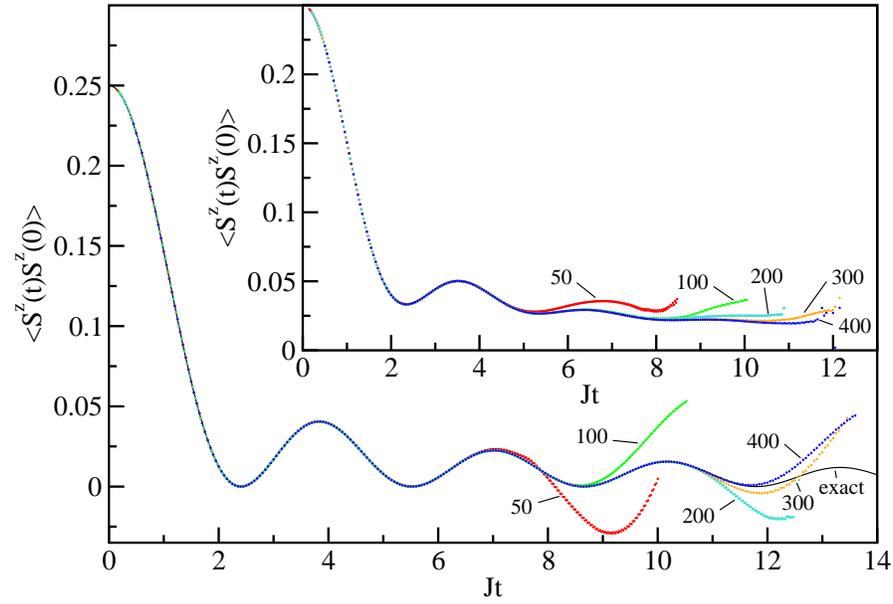}
\end{center}
\caption{Autocorrelation function for $\Delta=0$ and $\Delta=1$ (inset) at
  $T=\infty$ where $N=50-400$ states have been kept and $\delta=0.1$. The
  exact result is shown for comparison in the case $\Delta=0$.}
\label{Autocorr}
\end{figure}

\begin{figure}[!htp]
\begin{center}
\includegraphics*[width=0.99\columnwidth]{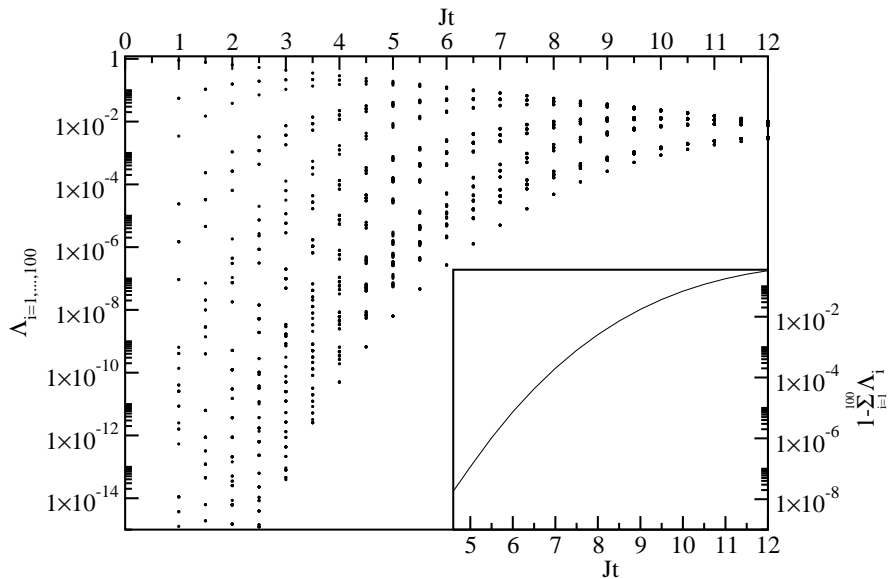}
\end{center}
\caption{Largest 100 eigenvalues $\Lambda_i$ of $\rho_S$
  for $\Delta=0$ and $T=\infty$ calculated exactly. The inset shows the
  discarded weight $1-\sum_{i=1}^{100} \Lambda_i$.}
\label{Spectrum}
\end{figure}

To demonstrate the method, results for the longitudinal spin-spin
autocorrelation function of the $XXZ$-chain at infinite temperature are shown
in Fig.~\ref{Autocorr}.
For $\Delta=0$ the $XXZ$-model corresponds to free spinless fermions and is
exactly solvable. We focus on the case of free fermions, as here the analysis
of the dynamical TMRG (DTMRG) method, its results and numerical errors can be
done to much greater extent than in the general case. The performance of the
DTMRG itself is expected to be independent of the strength of the interaction.
The comparison with the exact result in Fig.~\ref{Autocorr} shows that the
maximum time before the DTMRG algorithm breaks down increases with the number
of states. However, the improvement when taking $N=400$ instead of $N=300$
states is marginal. The reason for the breakdown of the DTMRG computation can
be traced back to an increase of the discarded weight (see inset of
Fig.~\ref{Spectrum}).
Throughout the RG procedure we keep only $N$ of the leading eigenstates of the
reduced density matrix $\rho_S$. As long as the discarded states carry a total
weight less than, say, $10^{-3}$ the results are faithful. For infinite
temperature and $\Delta=0$ we could explain the rapid increase of the
discarded weight with time by deriving an explicit expression for the leading
eigenstate of the QTM as well as for the corresponding reduced density matrix.
At the free fermion point the spectrum of this density matrix is
multiplicative. Hence, from the one-particle spectrum which is calculated by
simple numerics we obtain the entire spectrum. As shown in Fig.~\ref{Spectrum}
this spectrum becomes more dense with increasing time thus setting a
characteristic time scale $t_c(N)$, quite independent of the discretization
$\delta$ of the real time, where the algorithm breaks down. Despite these
limitations, it is often possible to extrapolate the numerical data to larger
times using physical arguments thus allowing to obtain frequency-dependent
quantities by a direct Fourier transform. This way the spin-lattice relaxation
rate for the Heisenberg chain has been successfully calculated
\cite{SirkerDiff}.

\acknowledgement S.G. acknowledges support by the DFG under Contracts No.
KL645/4-2 and J.S. by the DFG and NSERC. The numerical calculations have been
performed in part using the Westgrid Facility (Canada).

%\suPart \cite{Sirker}bsection{Subsection Heading}
%\label{sec:2}
%Your text goes here.

%\begin{equation}
%\vec{a}\times\vec{b}=\vec{c}
%\end{equation}

%\subsubsection{Subsubsection Heading}
%Your text goes here. Use the \LaTeX\ automatism for cross-references as
%well as for your citations, see Sect.~\ref{sec:1}.

%\paragraph{Paragraph Heading} %
%Your text goes here.

%\subparagraph{Subparagraph Heading.} Your text goes here.%
%
%\index{paragraph}
% Use the \index{} command to code your index words
%\begin{table}
%\centering
%\begin{tabular}{lll}
%\hline\noalign{\smallskip}
%\end{tabular}
%\end{table}
%\begin{figure}
%\centering
%% Use the relevant command for your figure-insertion program
%% to insert the figure file.
%% For example, with the option graphics use
%\includegraphics[height=4cm]{figure.eps}
%%
%% If not, use
%%\picplace{5cm}{2cm} % Give the correct figure height and width in cm
%\end{figure}
%
% For built-in environments use
%
%\begin{theorem}
%Theorem text\footnote{Footnote} goes here.
%\end{theorem}
%
% or
%
%\begin{lemma}
%Lemma text goes here.
%\end{lemma}
%
%
% BibTeX users please use
\bibliographystyle{spphys}
%% \bibliography{/home/js/TMRG/Greifswald_proceedings/ref_TMRG}
%% \bibliography{Literatur}

%
% Non-BibTeX users please follow the syntax
% the syntax of "referenc.tex" for your own citations
%%\input{referenc}
%%%%%%%%%%%%%%%%%%%%%%%%%%%%%%%%%%%%%%%%%%%%%%%%%%%%%%%%%%%%%%%%%%%%%%  }

%%%%%%%%%%%%%%%%%%%%%%%%%%%%%%%%%%%%%%%%%%%%%%%%%%%%%%%%%%%%%%%%%%%%%%

\printindex
\end{document}